\documentclass[final]{svjour3}
\usepackage{graphicx}
\usepackage{rotating}
\usepackage{amssymb}
\usepackage{mathptmx}
\usepackage[numbers]{natbib}
\usepackage{enumitem}
\makeatletter
\journalname{Journal of Low Temperature Physics}
\bibpunct{}{}{,}{s}{}{,}

\begin{document}

\title{A robust principal component analysis for outlier identification in messy microcalorimeter data}

\author{J.W.~Fowler   \and B.~K.~Alpert \and Y.-I.~Joe \and G.~C.~O'Neil \and D.~S.~Swetz \and J.~N.~Ullom}

\institute{Quantum Sensors Group, National Institute of Standards and Technology, 325 Broadway, Boulder, Colorado 80305, USA. \\
Fowler, Joe, and Ullom are also with University of Colorado, Department of Physics, Boulder, Colorado 80309, USA.  \\
\email{joe.fowler@nist.gov}}

\vspace{-1cm}
 \date{Draft of: \today}
\maketitle
\vspace{-5mm}

\begin{abstract}

A principal component analysis (PCA) of clean microcalorimeter pulse records can be a first step beyond statistically optimal linear filtering of pulses towards a fully nonlinear analysis.  For PCA to be practical on spectrometers with hundreds of sensors, an automated identification of clean pulses is required. Robust forms of PCA are the subject of active research in machine learning. We examine a version known as coherence pursuit that is simple, fast, and well matched to the automatic identification of outlier records, as needed for microcalorimeter pulse analysis.


\end{abstract}

\newcommand{\mat}[1]{\ensuremath{\mathbf #1}}   
\newcommand{\matT}[1]{\ensuremath{\mathbf #1}^\mathrm{T}}   
\newcommand{\inv}[1]{\ensuremath{{\mathbf #1}^{-1}}}   
\renewcommand{\vec}[1]{\mat{#1}}
\newcommand{\vecT}[1]{\matT{#1}}
\newcommand{\unitvec}[1]{\ensuremath{\hat{\vec{#1}}}}

\newcommand{\dif}    {\ensuremath{\,\mathrm{d}}} 

\newcommand{\ave}[1]{\left\langle#1\right\rangle}

\newcommand{\boxedResult}[1]{\fbox{$\displaystyle #1$}}

\newcommand{\me}{\mathrm{e}}
\newcommand{\mi}{\mathrm{i}}
\newcommand{\be}{\begin{equation}}
\newcommand{\ee}{\end{equation}}
\newcommand{\ba}{\begin{align}}
\newcommand{\ea}{\end{align}}

\section{Introduction}

The analysis of microcalorimeter pulses generally employs statistically optimal linear filtering.\cite{mosely1984,symkowiak1993,alpert2013} This approach starts from the assumption that all pulses have the same shape, so an optimal estimation of their amplitude serves as a monotonically increasing indication of the photon energy deposited in the absorber. As transition-edge sensors (TES) are sometimes operated near to their saturation energy, this assumption can fail badly enough that so-called ``optimal filtering'' becomes clearly sub-optimal. Also, nonlinear analysis of TES pulse data has been proposed as a way to improve energy linearity.\cite{bandler2006,peille2016,pappas2018} There is no clear consensus among low-temperature microcalorimeter researchers about how we should analyze data beyond the linear, single-shape assumption. Published proposals include: linear interpolation among a set of energy-specific pulse templates,\cite{shank2014} Taylor expansion of a continuous pulse shape model to leading order in energy,\cite{fowlertangent2017} and local linearization of a nonlinear manifold.\cite{fixsen2014,fowler2018}

A logical first step into nonlinear analysis is the projection of pulse records into a low-dimensional subspace. Such a projection is fully linear and can be made noise-optimal in the same sense as traditional optimal filtering. If the subspace is properly selected, projection would preserve most of the signal effects---linear or not---yet eliminate much of the noise. A nonlinear analysis could then begin from the projection, which would naturally be of much lower dimension than a complete pulse record: we find that the subspace typically requires no more than six dimensions, far fewer than the hundreds or thousands of samples that characterize a typical microcalorimeter pulse in its raw form. 

Another advantage of subspace projection lies in the separation of good pulses from statistical outliers. This advantage is relevant even to highly linear microcalorimeters such as metallic magnetic calorimeters.\cite{kempf2018} The residual pulse after projection should be quite sensitive to whether a pulse does or does not conform to the model implied by the subspace. Failure to conform produces large residuals.
The advantages of projection onto a low-dimensional subspace are thus at least threefold: dimensionality reduction; removal of a large share of the noise; and identification of outlier records.

A principal component analysis (PCA) of representative, clean pulses is one way to identify a productive subspace.~\cite{busch2016,yan2016} A normal data stream, however, is a mixture of clean pulses and unwanted records. Unwanted records include those containing two or more piled-up pulses or the decaying tails of earlier pulses; they may also include pathologies of the readout system that are difficult to model a priori. While the analysis of certain pulse-summary quantities can undoubtedly allow one to identify and eliminate these unwanted records, such analysis can be cumbersome and require unreasonable amounts of attention from human experts if applied to arrays of hundreds of sensors. An identical problem plagues machine learning, where there is often great value in a low-dimensional model that can describe most but not all data examples.\cite{candes2011} We have examined several robust variations on PCA, in search of one that suits the needs of microcalorimeter pulse analysis.

\section{Principal component analysis}

Principal component analysis finds the best rank-$r$ approximation to a matrix $\mat{M}\in\mathbb{R}^{m\times n}$ of dimensions much larger than $r$. Here, ``best'' means the smallest sum of square residuals, among other properties.\cite{golub} That is, it solves:
\be
\mathrm{minimize}\  || \mat{M}-\mat{U}\matT{U}\mat{M}||_\mathrm{F},
\ \mathrm{with\ respect\ to}\ {\mat{U}},
\ \ \ \mathrm{subject\ to\ } \matT{U}\mat{U}=\mat{I},
\ee
where $\mat{U}\in\mathbb{R}^{m\times r}$, $r<m \le n$, and $|| \cdot ||_\mathrm{F}$ is the Frobenius norm of a matrix. The matrix \mat{U} is an orthonormal basis for a subspace of dimension $r$. PCA guarantees that the subspace that \mat{U} spans best represents the columns of \mat{M}.
The principal components can be found by singular-value decomposition (SVD) of \mat{M} or by eigenvalue decomposition of $\mat{M}\matT{M}$.
Specifically, we can compute the PCA via the SVD as $\mat{M} = \mat{U}\mat{S}\matT{V}$, where the squares of the elements of the diagonal matrix $\mat{S}$ (the squares of the singular values) are the amount of the variance in \mat{M} accounted for by each principal component, and the columns of \mat{U} corresponding to the largest singular values are the leading principal components.

For purposes of pulse analysis, let the matrix $\mat{M}\in\mathbb{R}^{m\times n}$ contain thousands of pulse records all from the same sensor, with each of the $n$ columns representing one pulse record of length $m$. These records of ``training data'' should span the range of photon energies of interest, so that the principal components of \mat{M} will represent all future clean, single-pulse records over the broadest possible energy range.  PCA analysis generally starts with recentering and rescaling each column of \mat{M}, subtracting its mean and dividing by its standard deviation to achieve zero mean and unit variance.  In this paper, we make a different, related adjustment. A global constant (the median value of all pretrigger samples) is subtracted from \mat{M}, because DC offsets are both meaningless and large in typical microcalorimeter data.   

Suppose that the matrix \mat{M} is dominated by signal, as it is with the very high-resolution pulses of an x-ray microcalorimeter. Then the leading principal components of \mat{M} will define, among all possible data vectors, a low-dimensional subspace that approximately contains all actual pulses. Projection of raw pulse records into this subspace can serve the dual purposes of compressing the pulse records into a few, most informative numbers apiece and eliminating a large fraction of the noise. A subspace of higher dimension will capture both more signal and more noise than will one of lower dimension.

If the goal is identification of the appropriate subspace for clean pulses, then we actually require not the full PCA or SVD, but only the column space of the matrix, and in fact, only those columns corresponding to the few largest singular values. Algorithms faster than the full SVD can be used to determine leading singular vectors. The \emph{truncated SVD}\cite{larsen1998} is an iterative algorithm to compute singular vectors in order, implemented in the PROPACK library.\cite{propack} Also fast and very simple to implement are \emph{randomized SVD} algorithms.\cite{halko2011} They can compute an excellent approximation to the column space of a matrix and the leading singular values, and they are well-suited to the problem of finding a low-dimensional subspace that contains most of the signal in pulse records.

\section{Robust methods of PCA}

\begin{figure}[tbp]
\begin{center}
\includegraphics[width=\linewidth, keepaspectratio]{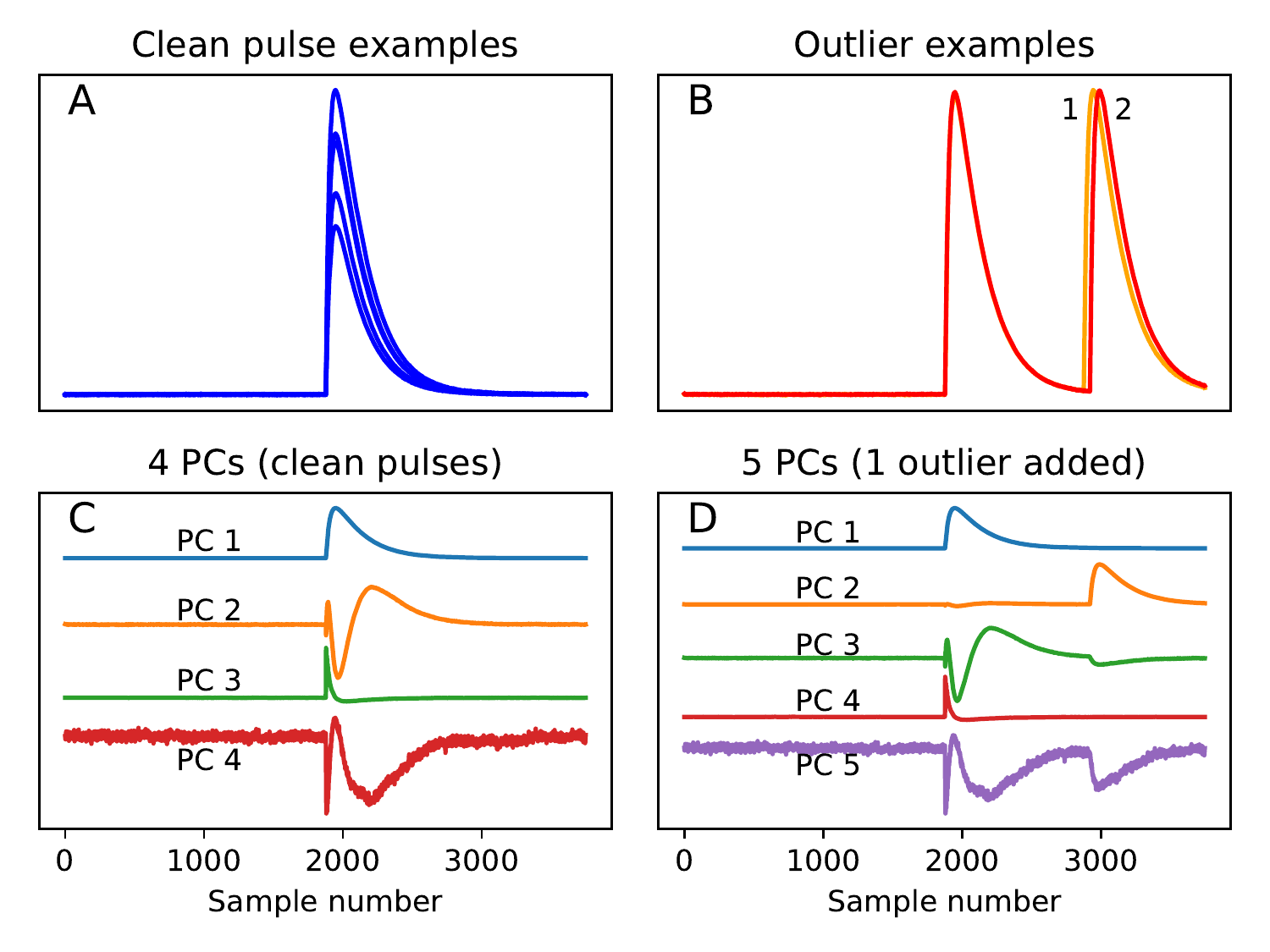}
\caption{\label{fig:examples}
Example pulses. {\it Top left:} several clean pulses of various energies. 
{\it Top right:} two examples of pileup, which we wish to flag as outliers.
{\it Bottom left:} four principal components determined by SVD of a set of 1500 clean pulses. Crudely, the first represents a pulse; the second, the difference between pulses of high and lower energy; and the third, the difference between pulses that arrive early or late with respect to the sampling clock. Further singular components lack simple explanations.
{\it Bottom right:} five principal components from a ``contaminated'' set: the same 1500 clean pulses plus the single outlier marked {\bf 2} in the top right examples. Erroneous inclusion of this single outlier  increases the dimension of the needed subspace by one. Also, it undermines the rejection of future outliers based on root-mean-square (rms) difference between pulse records and their projection into the low-dimensional subspace. For example, the typical rms residual of normalized clean pulses is 0.006 after projection into either basis. The outliers 1 and 2, however, have rms residual of 0.66 after projection into the clean basis but only 0.29 and 0.001 after projection into the contaminated basis. 
(Color figure online.)}
\end{center}
\end{figure}

Unfortunately, the presence of outliers in the training data \mat{M} can be disastrous. Even a single column out of thousands that contains, for example, two piled-up pulses can easily affect the first several principal components. This distorts our estimate of the space spanned by all normal, single pulses with two deleterious effects. First, we are effectively wasting one of the few numbers intended as a summary of clean pulses, using it to carry unwanted information instead. Second, we would like to use the residual after projection into the subspace as a test for outlier pulses. The residual should be large if and only if a record looks nothing like a pulse, whereas piled-up double pulses with timing and relative amplitude similar to an outlier will have reduced residuals if that outlier is included in the PCA, as Figure~\ref{fig:examples} shows. What is needed is a variation on PCA that is immune to outliers in the training data.

The search for robust methods of PCA is a very active area of research in statistics and machine learning. \cite{bouwmans2014} This search is sometimes divided into \emph{robust PCA}, where individual matrix elements of \mat{M} are outliers,\cite{vaswani2018} and \emph{robust subspace recovery}, where entire columns may be outliers.\cite{lerman2018}
We have considered \emph{L1-norm PCA},\cite{kwak2008,nie2011,park2016,markopoulos2017}
 in which principal components minimize not the variance of the data residuals (the L2 norm) but the absolute deviation of the residuals (the L1 norm). While this PCA is more tolerant of a few pathological pulse records, we find it not useful for the current application: it tends to reduce rather than eliminate the influence of outliers. The same is true of L2,1-norm PCA,\cite{nie2016} which addresses a certain practical objection to L1-norm PCA.

Two other algorithms for robust subspace recovery operate by cleaning the data matrix \mat{M} through the identification and removal of outlier columns, after which a standard PCA can be performed. The \emph{outlier pursuit} method\cite{xu2012} is more productive than the L1- and L2,1-norm versions of PCA for the data-cleaning problem, but it is relatively slow and depends on the user setting two separate free parameters that might be difficult to determine automatically.

The method best suited to microcalorimeter analysis is known as \emph{coherence pursuit}.\cite{rahmani2016} It relies on the idea that good columns in \mat{M} (here, clean pulses) will tend to lie in nearly the same direction in $\mathbb{R}^m$ as many other columns, while outlier columns will tend to lie far from all or 
\renewcommand{\thefootnote}{\ensuremath{\fnsymbol{footnote}}}
most other columns, even in the extreme case that outliers outnumber the clean pulses.\footnote{Outlier pursuit thus relies on the \emph{Anna Karenina} principle: clean pulse records are all alike; every unclean record is unclean in its own way.}
\renewcommand{\thefootnote}{\arabic{footnote}}
The underlying idea that random high-dimensional points are nearly always orthogonal to one another can be made precise.~\cite{cai2013}
The \emph{mutual coherence} of two pulses is defined as the absolute value of the cosine of the angle between the column vectors. When we compute the mutual coherence for all pairs of pulse records, outliers tend to have a much lower sum of mutual coherence with all other records than non-outliers do. Coherence pursuit has a major advantage over most other robust PCA methods in that it is non-iterative; most competing methods require numerous iterations with expensive steps in each, such as an SVD.

Specifically, the coherence pursuit algorithm applied to pulse records is:
\begin{enumerate}[topsep=1mm]
\item Subtract the median pretrigger value $d$ from all samples: $\mat{M}_0\equiv\mat{M}-d$. Thus further steps are insensitive to a meaningless and arbitrarily large overall offset, but the variations in baseline level from one record to another are preserved.
\item Compute the L2 norm \vec{r} of columns of $\mat{M}_0$, the \emph{pulse rms amplitude}: $r_j = [\sum_i\,(M_0)_{ij}^2]^{1/2}$.
\item Normalize the columns of $\mat{M}_0$ by creating the diagonal matrix \mat{R} from the pulse rms values and defining $\mat{X}\equiv \mat{M}_0 \inv{R}$.
\item Compute the pairwise mutual coherence matrix $\mat{G}=\matT{X}\mat{X}$.
\item Compute the coherence vector \vec{g} as the L1-norm of each column of \mat{G} (omitting the $g_{ii}=1$ diagonals). That is, $g_j = \sum_{i=1, i\ne j}^n\,|G_{ij}|$, as there are $n$ columns (pulses) in \mat{M}.
\item Define the \emph{mean coherence} as $\vec{c}\equiv\vec{g}/(n-1)$.  Given this scaling, values $c=0$ or 1 indicate a pulse record orthogonal to or parallel to all others, respectively.
\end{enumerate}
Pulse records $j$ with large values of $c_j$ are clean pulses; low values of $c_j$ indicate that record $j$ is an outlier. But where shall we draw the line? For nonlinear detectors, the answer to this question turns out to depend on photon energy. We return to this matter in the next section. Certain choices in the algorithm deserve further study on a wide range of data sets. In step 5, for example, we find no clear preference for the L1-norm over the L2-norm. Similarly, the median coherence appears to be an acceptable alternative to the mean in step 6. 

A related, non-iterative method for outlier removal\cite{menon2019}  also starts from the mutual coherence matrix \mat{G} and then rejects unstructured and structured (clustered) outliers in successive steps. It requires no free parameters. This attractive feature appears to work well only when outliers are uniformly random in $\mathbb{R}^m$, however. Outlier pulse records are often similar to clean pulses for much of their length, on the other hand, so they violate the core assumption that would allow the rejection criterion to be computed from first principles.

If the number of pulses to be analyzed, $n$, is large and it is  difficult to compute the full $\mat{G}\in\mathbb{R}^{n\times n}$ matrix, then we can subdivide the data into smaller batches and perform the outlier rejection step on each batch separately.
After outliers have been identified and removed from $\mat{M}_0$,
we can complete the robust PCA by computing the SVD of the cleaned matrix. 

\section{Coherence pursuit demonstrated for TES pulse records and the threshold question}\label{sec:demo}

\begin{figure}[htbp]
\begin{center}
\includegraphics[width=\linewidth, keepaspectratio]{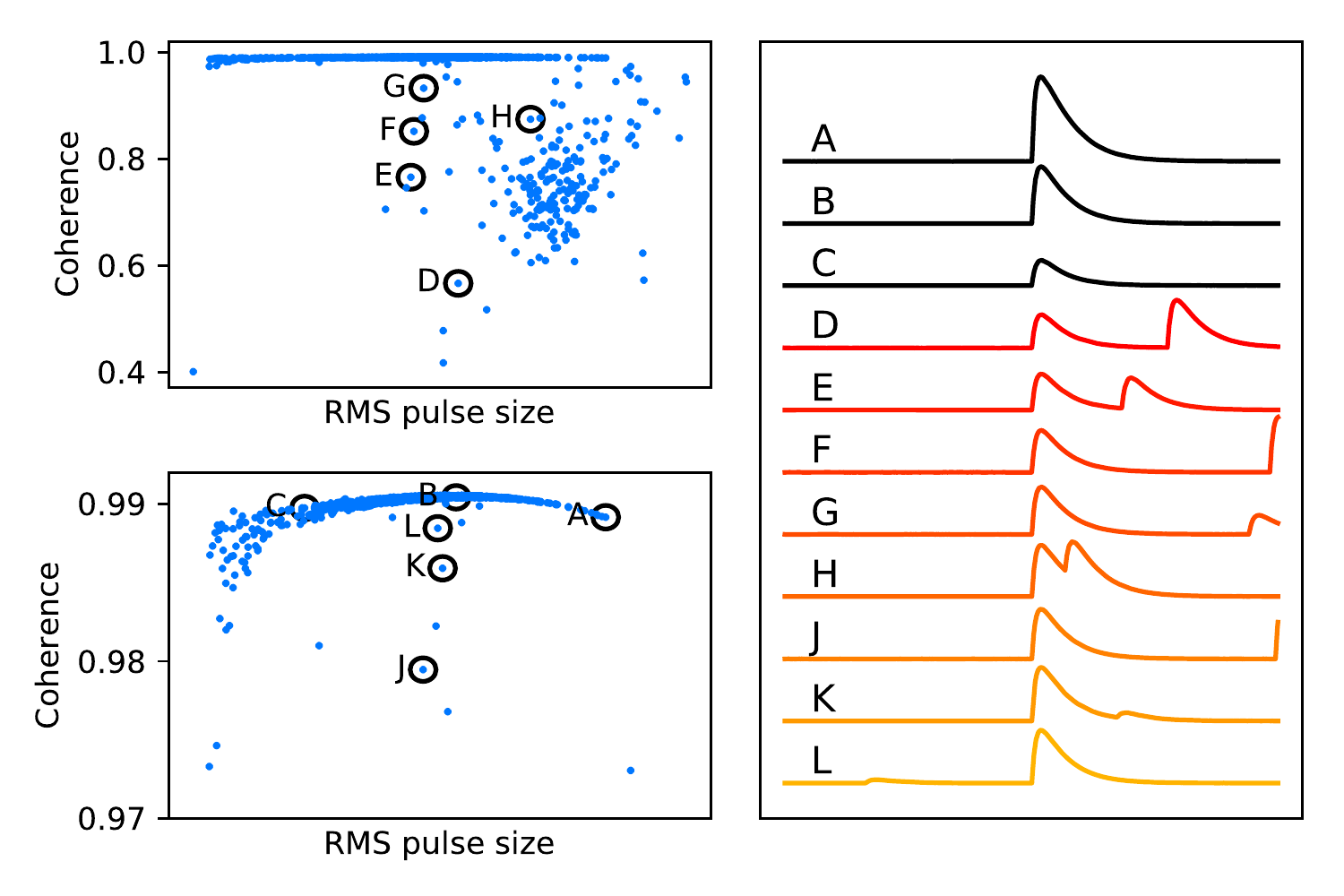}
\caption{\label{fig:coherence_vs_E}
{\it Left:} Two views (different $y$ ranges) of the mean absolute coherence $c$ as a function of rms pulse size, which in the absence of pileup serves as a rough estimator of photon energy (approximately 0 to 10\,keV is shown across the full $x$ range). The coherence of clean pulses is slightly less at the energy extremes than in the center of the spectrum.
{\it Right:} Eleven specific pulses, whose coherence and size are indicated by circles on the left-hand panels. A, B, C are clean pulses of large, medium, and small size. The others are identified as outliers by the coherence metric. D--H are clearly outliers, with $C<0.95$. Even J, K, and L contain small amounts of pileup, though $C>0.98$ for these examples.
(Color figure online.)}
\end{center}

\end{figure}

We demonstrate coherence pursuit on approximately 6000 pulse records from one sensor in a TES array in Figure~\ref{fig:coherence_vs_E}. The x-rays in the range 4\,keV to 10\,keV are the K-shell fluorescence lines of various transition metals and L-shell lines of certain lanthanide metals, similar to the data previously used for metrology studies.\cite{fowler2017} Records are selected arbitrarily but from all portions of a measurement lasting fifteen hours. Specific pulses are shown, including three clean pulses and eight outliers,  some of which differ only subtly from clean records.

The lower left panel of Figure~\ref{fig:coherence_vs_E} shows that good pulses have a coherence that depends on pulse amplitude. Clean pulses at the energy extremes, either high or low, show slightly lower coherence than clean pulses at the middle of the energy range. TES nonlinearity causes this effect: because pulse shapes change slightly with energy, pulses at the lowest or highest energies are less coherent with the ensemble of good pulses overall. While the effect is small, a PCA that fully accounts for pulse shapes at the energy extremes requires a threshold on the coherence \vec{c} that depends on pulse size or energy.

The entire robust PCA procedure can be no more automatic than our ability to select this coherence threshold curve. The following is a first attempt at such selection. We start from the notion that outliers are less coherent than good pulses, but no pulse can be \emph{more} coherent. Therefore, we can approximate the shape of the $c$ vs size curve by studying its maximum value vs size. A simple approach is to build a linear interpolation between some ten to twenty anchoring points. Starting at the point with the highest value of $c$, we require that the slope of the interpolation is a monotone decreasing function of rms pulse size. In this data set, any pulse with coherence $c$ differing from this model of the upper limit by at least $5\times10^{-3}$ can be considered an outlier. A key question about this method is whether this threshold is generally valid, and if not, whether it can be selected automatically. Our investigations so far suggest that one threshold should suffice for many measurements, so long as the experimental conditions (including spectra) are similar.

\section{Noise-weighted PCA and projections}

We have so far considered the standard version of PCA, which is statistically optimal only when signals contain exactly white noise. With microcalorimeter data, we generally have higher noise at low frequencies, below the inverse of the thermal and electrical time constants.\cite{irwin_hilton} If we can estimate \mat{R}, the noise covariance matrix in some way (perhaps by recording pulse-free data), then any matrix \mat{W} satisfying $\mat{W}\mat{R}\matT{W}=\mat{I}$ is a linear noise-whitening transformation.  That is, if raw data \vec{d} has covariance $\mathrm{E}(\vec{d}\vecT{d}) - \mathrm{E}(\vec{d})\mathrm{E}(\vecT{d}) = \mat{R}$, then whitened data $\mat{W}\vec{d}$ will have a covariance of \mat{I}, or white noise (of unit variance).\cite{fowler_mpf} One possible construction of \mat{W} is the inverse of the lower Cholesky factor of \mat{R}. In cases where the white noise assumption of robust PCA is inadequate, then it can performed on a noise-whitened version of the training data, on $\mat{W}\mat{M}$ instead of on \mat{M}. 
Coherence pursuit can used to remove outlier columns of either \mat{M} or $\mat{W}\mat{M}$ before performing this \emph{noise-weighted PCA}.

We do not have a clear sense of when or whether it is important to perform a noise-weighted PCA in microcalorimeter data analysis. With enough clean pulses in the training data set, one might always find principal components and noise-weighted ones spanning so nearly the same space as to be indistinguishable. We intend to compare the two versions in a range of future analyses, hoping to determine the value added by the noise-whitening step.

Whether PCA is noise-weighted or not, it is without a doubt important to perform \emph{projections} into the low-dimensional subspace with noise weighting. This step makes subspace projection optimal in the same sense as traditional optimal filtering is. Let \mat{U} be the approximate column space found by robust PCA, where \mat{U} has orthonormal columns ($\matT{U}\mat{U}=\mat{I}$). Then the optimal projection $\vec{d}_u$ of a pulse record \vec{d} into the subspace spanned by \mat{U} is:
\begin{eqnarray}
\vec{d}_u &=& \mat{U}(\matT{U}\inv{R}\mat{U})^{-1} \matT{U}\inv{R}\vec{d} \label{eq:proj}\end{eqnarray}
This is a projection in the sense that it is linear in \vec{d} and that
\be
\vec{d}_u = \mat{U}\matT{U}\vec{d}_u.
\ee
It is optimal with respect to noise in the sense that it is the projection that maximizes the likelihood of the observed data vector \vec{d} under the multivariate Gaussian noise model \mat{R}. We distinguish the projected pulse record $\vec{d}_u$ from the vector of principal component amplitudes \vec{p}, which one would use to estimate filtered pulse heights. The first is an approximation to the measured data; the second is a summary of it. They are related by  $\vec{p}=\matT{U}\vec{d}_u$ or $\vec{d}_u=\mat{U}\vec{p}$.  A geometric view of this optimality is that projection Eq.~\ref{eq:proj} minimizes the Mahalanobis distance between \vec{d} and its projection, $||\vec{d} - \vec{d}_u||_\mathrm{M}$, instead of the usual Euclidean distance $||\vec{d}-\vec{d}_u||$. The Mahalanobis, or signal-to-noise, norm is defined\cite{mahalanobis} as $||\vec{v}||_\mathrm{M} \equiv (\vecT{v}\inv{R}\vec{v})^{1/2}$.

\section{Conclusions}

Analysis of data from large arrays of microcalorimeters will require ever-increasing levels of automation in our pulse processing steps. If we are to entertain nonlinear analysis of pulses, we are likely to start with the data compression step of noise-optimal projection of pulse records into small linear subspaces revealed through PCA. We have described a robust, outlier-resistant form of PCA that can be applied to microcalorimeter pulses with minimal expert human intervention. The method is based on coherence pursuit to reject outliers from training data; randomized PCA of the cleaned data to find an appropriate subspace; and noise-optimal projection of pulse records into this subspace. 

The problem we have addressed here is only one of many hard and unsolved problems in high-resolution pulse analysis. Other examples include learning detector parameters and the governing electro-thermal differential equations from data, modeling detector noise from noisy data, reducing the effects of crosstalk, and zero-prior-knowledge alignment of uncalibrated spectra. One feature that many of these hard problems share is that they are unlikely to have simple, analytic, and clean solutions. Another is that solutions must be automated for practical use with large calorimeter arrays. In both senses, they have much in common with the problems faced in machine learning. We believe that the enormous attention now being turned upon machine learning problems, and the vast number of creative ideas being generated will continue to benefit our work on microcalorimeter data in the future.

\begin{acknowledgements}
This work was supported by NIST's Innovations in Measurement Science program and by NASA SAT NNG16PT18I, ``Enabling \& enhancing technologies for a demonstration model of the Athena X-IFU.'' We thank Dan Becker and Malcolm Durkin for numerous discussions and earlier work on pulse outlier identification.
\end{acknowledgements}


\end{document}